\documentclass[a4paper,11pt]{article}
\usepackage[cp1251]{inputenc}%for Windows cyrillic encoding
\usepackage[russian]{babel}
\usepackage{graphicx}
\usepackage{amsmath}
\usepackage{amssymb}
\usepackage{amstext}
\begin{document}
\renewcommand{\figurename}{Figure}% for text in English
\renewcommand{\tablename}{Table}% for text in English

%\bigskip
\centerline{\textbf{Comparison of the infrared magnitudes of the eclipsing variable star $\varepsilon$ Aur}}
\centerline{\textbf{during the two main luminance lows using observations from 1980 to 2015}}
\bigskip
\centerline{\textbf{I.A.~Maslov$^{1,2}$, O.G.~Taranova$^2$, A.M.~Tatarnikov$^2$, V.I.~Shenavrin$^2$}}
\smallskip
\centerline{\textit{$^1$Space Research Institute (IKI), 117997, 84/32 Profsoyuznaya Str, Moscow, Russia}}
\centerline{\textit{E-mail: imaslov@iki.rssi.ru}}
\centerline{\textit{$^2$Moscow State University, Sternberg astronomical Institute, Universitetsky pr., 13, Moscow 119991,Russia}}
%\centerline{\textit{$^3$Author1's address and affiliation}}
\medskip
%\centerline{\small Received August 31, 2015}
\bigskip
\textbf{Abstract.}
The data of photometric measurements of the long-period eclipsing
variable $\varepsilon$ Aur in the two main minima 1982 and 2010 in the spectral range of 1--5 microns are presented.
Noted is the similarity of light curves in eclipse, its asymmetry, and availability of the
short interval increased brightness.
We are detecting the phase changing of the indicator color J-M and the effect of short-time 
changing of the color to be more blue  at the moments during the beginning and end of the eclipse.


\section*{Introduction}
Eclipsing system $\varepsilon$ Aur (HD~31964) consist of a class F Supergiant
orbiting, with a period 27.1\,year, a more massive and invisible hot component, surrounded
by a disk of dust particles.
A reliable parallax and orbital parameters were determined by Kamp (1978) on the basis of astrometric data.
The primary minimum occurs when partially shading the Supergiant from a disk, 
this was confirmed by interferometric observations of  Kloppenborg (2010).
There are work of Maslov (2014) indicates the presence of a secondary minimum, 
when the Supergiant  is covering  the central part of the disk, which is heated by the massive component.
Spectrophotometric analysis of Taranova (2001) and Hoard (2010) shows the presence of an infrared excess in the spectrum, 
which indicates that the dust particles in the disk are heated.
Budai (2011) proposed that the disk consists of a large-size dust particles with strong forward scattering for short wavelength. 

\section*{Observations and Results}

Observations of Epsilon Auriga in the infrared region of the spectrum was carried out by us since 1980 
at the 1.25-cm telescope of the South Laboratory of the Sternberg astronomical Institute.
Until 1985 it was used a two-channel photometer on the bands of 1.2--2.2 microns based on PbS-detector; see Nadzhip (1986).
After 1985, the observations were made in the five bands from 1.2 to 4.8 micrometers 
using a modulation mode photometer based on the liquid nitrogen cooled InSb-photodiode; see Moroz (1979).
Photometric calibration was performed by observing the stars BS1454.
Figure~1 shows the light curves in the band J near main lows 1982 and 2010.
There is a similarity of light curves in both eclipses, asymmetry and availability of the
short interval increased brightness at orbital phase 1.05.
Figure~2 shows the dependence of the color index J-M from the phase of the orbital period.
For clarity, we averaged the data of our measurements in groups of 3 to 13 nights; values of the average and its error are represented.
For comparison, figure~3 shows a similar dependence of the distance from the Supergiant to the disk .

\begin{figure}[H]
\begin{center}
\hspace*{-2cm}
\includegraphics*[height=.75\textheight,angle=270]{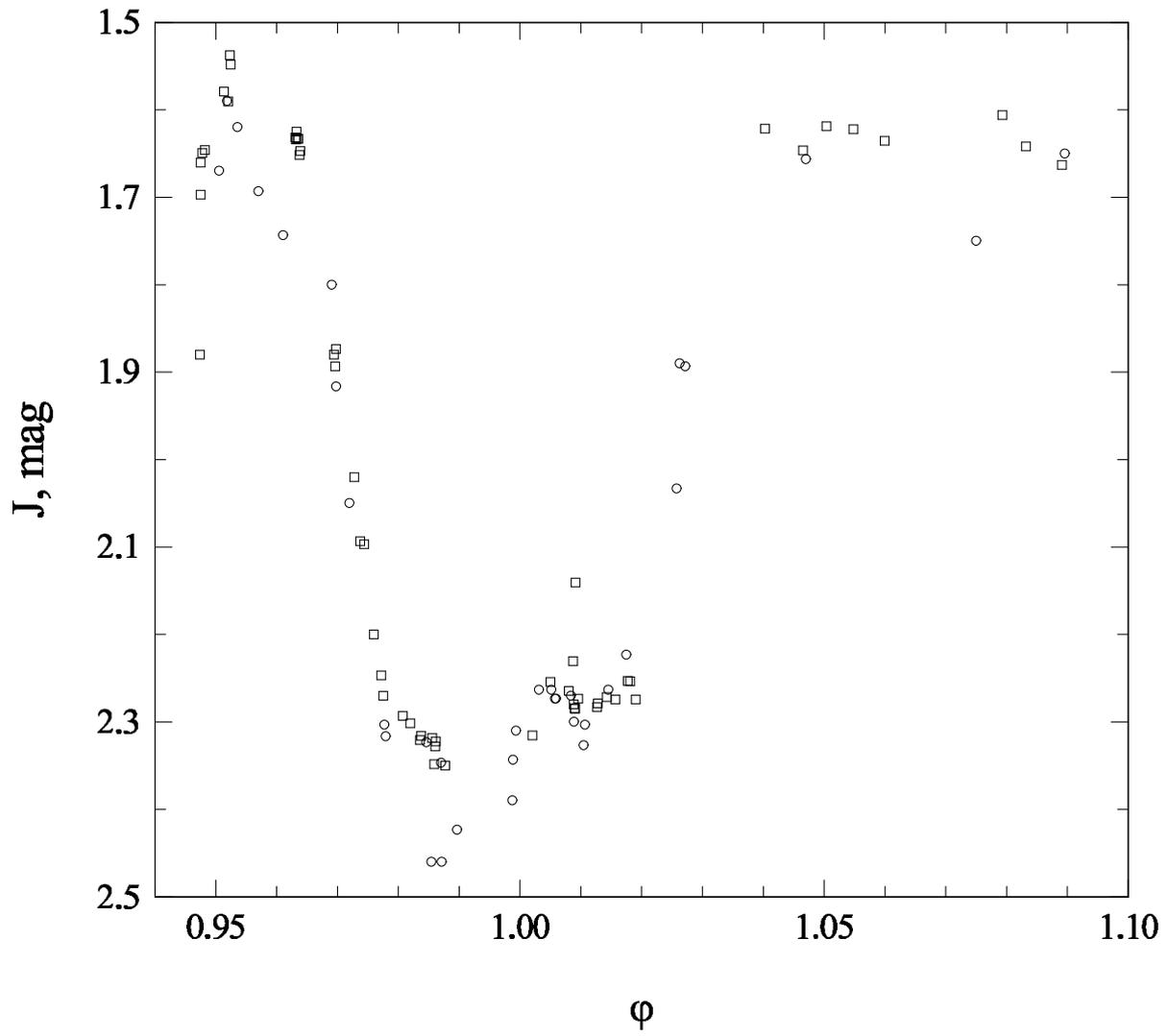}
\end{center}
\caption{J-values depending on the phase of the brightness 
near the lows of 1982 (circles) and 2010 (squares).}
\end{figure}

\begin{figure}[H]
\begin{center}
\hspace{-10mm}
\includegraphics*[height=.53\textheight,angle=270]{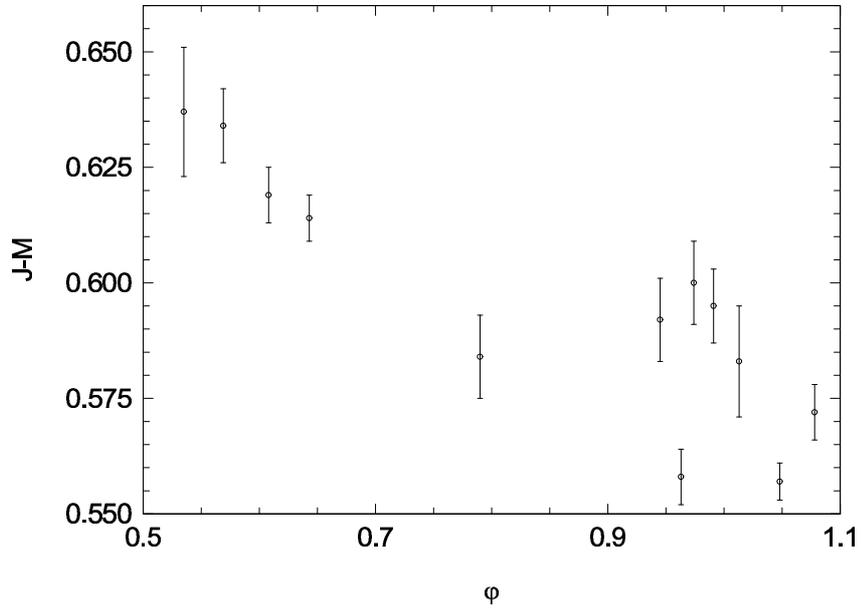}
\end{center}
\caption{The dependence of the color index J-M from the orbital phase.}
\end{figure}

\begin{figure}[H]
\begin{center}
\hspace*{-6mm}
\includegraphics*[height=.5\textheight,angle=270]{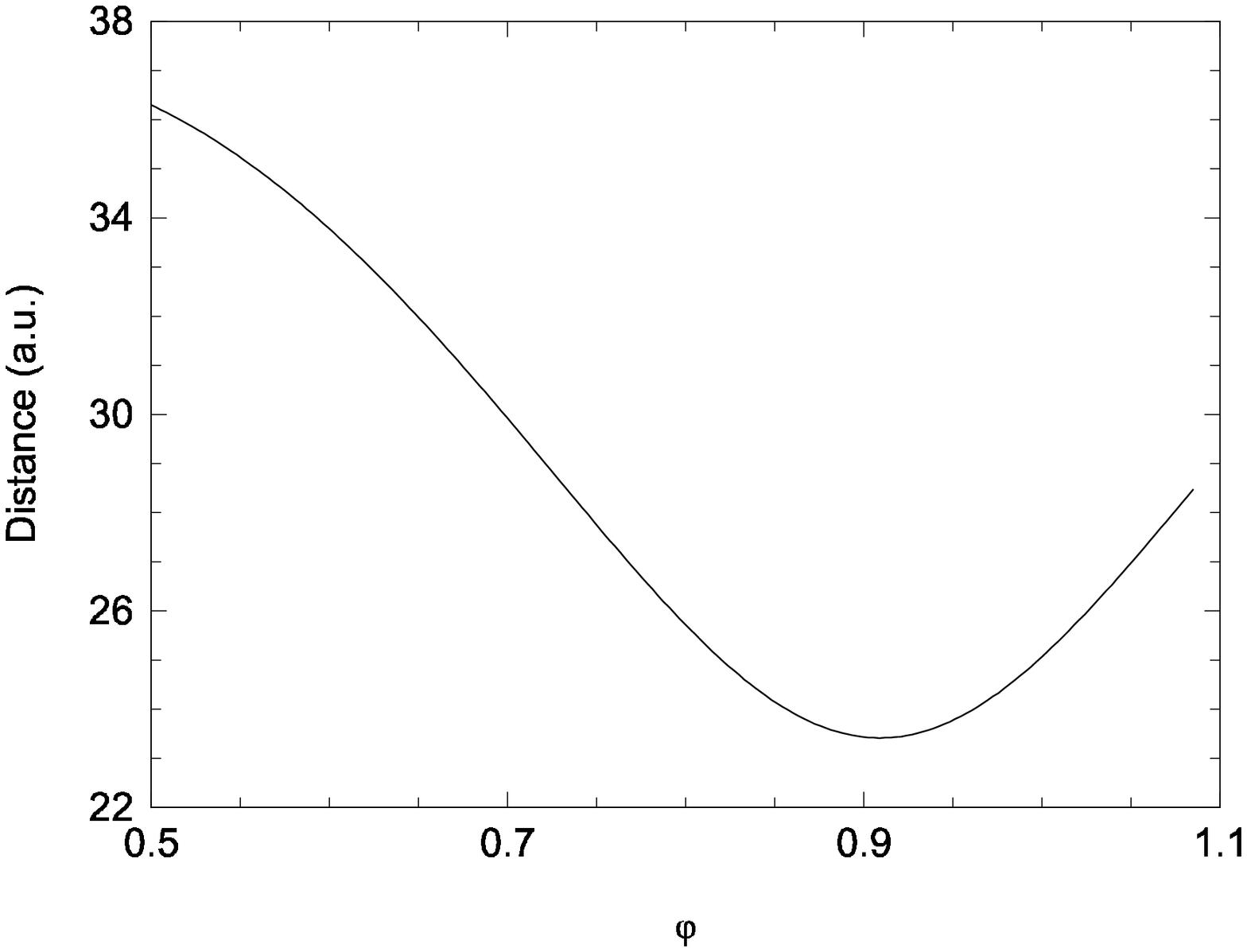}
\end{center}
\caption{The dependence of the distance (a.u.) between the stars of the system depending from the orbital phase.}
\end{figure}

\section*{Conclusion}

Light curves in the lows of 1982 and 2010  in the near infrared spectral region are similar to each other.
They are asymmetric near zero phase, which is probably due to the asymmetry of the disk.
Availability of the short interval increased brightness  near phase 1.05 is explained by the decreased
thickness of the disc near its central part, where exists the heavy component of the system.
The phase changing indicator color J-M can be explained by the fact that distance from disk to the F-star is changing.
As result, the temperature of the dust-particles are changing and it is changing its part in the common light flux.
In addition, we are detecting changing of the color to be more blue  at the moments of the beginning and end of the eclipse,
which we attribute to scattering of light at the edges of the dust disk. 
This is possible if the angular distribution of light scattering on dust 
significantly extended forward at a wavelength of 1 micron compared to a wavelength of 5 microns.

\section*{References}

\hspace{4.8mm}
 
Budaj~J. (2011) A\&A 532, L12.

Hoard~D.W., Howell~S.B., Stencel~R.E. (2010) AJ 714, 549.

Kamp~P. van de (1978) AJ 83, 975.

Kloppenborg~B., Stencel~R., Monnier~J.D. et. al. (2010) Nature 464, 870.

Maslov~I.A., Nadjip~A.E., Taranova~O.G., Tatarnikov~A.M., Shenavrin~V.I. (2014) Ap 57, 370.

Moroz~V.I., Taranova~O.G., Shenavrin~V.I., Yudin~B.F. (1979) ATsir 1056.

Nadzhip~A.E, Shenavrin~V.I., Tikhonov~V.G (1986) Tr. Gos. Astron. Inst., Mosk. Gos. Univ. 58, 119.

Taranova~O.G., Shenavrin~V.I. (2001) AstL 27, 338.

\end{document}